\documentclass[english,journal=jpccck,manuscript=article]{achemso}
\usepackage[T1]{fontenc}
\usepackage[latin9]{inputenc}
\usepackage{xcolor}
\usepackage{textcomp}
\usepackage{multirow}
\usepackage{amstext}
\usepackage{graphicx}
\usepackage{wasysym}
\usepackage{esint}

\makeatletter

\newcommand{\lyxmathsym}[1]{\ifmmode\begingroup\def\b@ld{bold}
  \text{\ifx\math@version\b@ld\bfseries\fi#1}\endgroup\else#1\fi}

\title{ScS$_{2}$ Monolayer as a Potential Cathode Material for Alkali-ion
Batteries and Beyond}

\author{Dwaipayan Chakraborty}

\email{dc167@snu.edu.in}

\affiliation{Department of Physics, School of Natural Sciences, Shiv Nadar University,
Greater Noida, Gautam Buddha Nagar, U.P. (India) -- 201314}

\author{Madhu Pandey}

\author{Priya Johari}

\email{priya.johari@snu.edu.in}

\affiliation{Department of Physics, School of Natural Sciences, Shiv Nadar University,
Greater Noida, Gautam Buddha Nagar, U.P. (India) -- 201314}

\providecommand{\tabularnewline}{\\}

\makeatother

\usepackage{babel}
\begin{document}

\subparagraph{2D materials, transition metal dichalcogenides, ScS$_{2}$ monolayer,
cathode materials, open circuit volatge gravimetric capacity, diffusion
kinetics}

\section{Introduction:}

With the decreasing stock of fossil fuels and increasing environmental
pollution, the demand for green energy sources and rechargeable energy
storage devices in recent times is like never before. Lithium-ion
batteries (LIBs) have been in the forefront of the secondary battery
technology since its first commercialization in 1991\cite{1}. LIBs
are known to deliver high energy density, high reversible capacity,
excellent cycling stability, and to have relatively simple reaction
mechanism due to which they are being widely used in portable electronic
devices and in recent time, in electric vehicles\cite{2,3}. However,
limited lithium resources in highly localized manner in few geographical
locations and the water extensive mining processes hinder the cost
effectiveness of LIBs and create environmental concerns. Along with
this there are also intrinsic safety issues with LIBs. All of these
drawbacks collectively push to search for other metal-ion batteries\cite{4,5}.
In this context sodium-ion batteries (NIBs), potassium-ion batteries
(KIBs), magnesium-ion batteries and aluminum-ion batteries have gained
huge research interests in recent times due to relatively higher abundance
of these metals in the earth's crust. 

Any metal-ion battery has main three components namely anode, electrolyte
and cathode among which cathode is the most expensive and has largest
weight, and thus, controls the cost and the specific capacity of the
battery to a large extent\cite{3}. Therefore, developing high energy
high power cathode materials for LIBs and non-LIBs is a key challenge.
However, LiCoO$_{2}$, the most popular cathode material in commercial
LIBs delivers specific capacity around 140 mA h g$^{-1}$ only\cite{6}.
Other layered oxide materials like LiMnO$_{2}$, LiNi$_{0.5}$Mn$_{0.5}$O$_{2}$
also struggle to cross 200 mA h g$^{-1}$ specific capacity\cite{3}.
On the other hand, the olivine LiFePO$_{4}$ has also been reported
to deliver the specific capacity $\sim$140 mA h g$^{-1}$ only with
better cycling stability\cite{7}. The same situation can be evidenced
for NIBs also. Layered cathode materials like NaMO$_{2}$ (M = Mn,
Ni, Co) and there derivatives can only deliver theoretical capacities
around 250 mA h g$^{-1}$\cite{8}. Polyanionic compounds like NaFePO$_{4}$,
Na$_{3}$V$_{2}$(PO$_{4}$)$_{3}$ deliver even less theoretical
capacity of 154 mA h g$^{-1}$ and 117 mA h g$^{-1}$, respectively\cite{8}.

Also the preparation of well-ordered bulk structures are difficult
and layered materials suffer from capacity fading in the bulk form\cite{9}.
In this context, due to relatively low weight, large surface to volume
ratio, good conductivity on the surface, mechanical stability etc.,
several 2D materials have gained huge scientific attention for their
exploration as new generation electrode materials. Although graphene
is not an ideal electrode for LIBs\cite{10} but other graphene-like
2D materials such as silicene\cite{11}, germanene\cite{12}, borophene\cite{13}
and phosphorene\cite{14,14a} have been shown to deliver high specific
capacity when used as anode. Numerous other 2D materials explored
mainly as anode like metal oxides and metal nitrides\cite{15}, transition-metal
dichalcogenides (TMDCs)\cite{16}, MXenes\cite{16} exhibit promising
electrochemical performances with large specific capacities, high
rate capabilities, and good cycling stabilities. On the cathode side
a few 2D materials like MoN$_{2}$\cite{2}, MnO$_{2}$, CoO$_{2}$,
NiO$_{2}$\cite{9}, ScO$_{2}$\cite{17}, V$_{2}$O$_{5}$ monolayer\cite{18},
NbS$_{2}$\cite{19} for alkali-ion batteries, MoS$_{2}$ nanoribbon
for Mg-ion batteries\cite{20} etc. have been theoretically studied
in the recent times. 

Inspired from these studies and the fact that Sc is the lightest transition
metal which could help to achieve the goal of high theoretical capacity,
we here explored the performance of ScS$_{2}$ monolayer as a cathode
material for alkali-ion batteries (Li, Na, K) and other multi-valent
metal-ion batteries (Mg, Al). Previous studies on ScS$_{2}$ have
focused only on the fundamental electronic and magnetic properties
of the ScS$_{2}$ monolayer, but not on it's possible applications\cite{21,22}.
Our first-principles calculations show that 2D ScS$_{2}$ is able
to deliver large theoretical capacity of 491.36 mA h g$^{-1}$ for
alkali-ions and 324.29 mA h g$^{-1}$ for Mg and Al-ions while maintaining
good average open-circuit voltages. We also studied the diffusivity
of these metal-ions on the ScS$_{2}$ surface which is related to
the charge/discharge rate capability of batteries. Our results suggest
low diffusion barriers for all metal-ions except Al. Owing to these
results, we therefore believe that the ScS$_{2}$ monolayer can be
an interesting candidate for cathode material to be used in alkali-ion
batteries and beyond.

\section{Computational Details:}

We employed the first-principles calculations in the framework of
the density functional theory (DFT) as implemented in the VASP (Vienna
ab initio simulation package)\cite{23,24,25} package to study the
performance of ScS$_{2}$ monolayer as cathode for metal-ion (Li,
Na, K, Mg, Al) batteries. We used the projector augmented-wave (PAW)
pseudopotentials\cite{26} to treat the electron-ion interactions
with a plane-wave cutoff energy of 400 eV. The electron-electron exchange
correlation was described by the generalized gradient approximation
(GGA) within the Perdew--Burke--Ernzerhof\cite{27} formalism. We
also incorporated the vdW effect in our adsorption calculations by
considering the DFT-D3 method of Grimme\cite{28} for empirical dispersion
corrections. The energy and the Hellmann\textminus Feynman force convergence
criteria of 10$^{-4}$ eV and 0.01 eV/Å, respectively were used for
all the calculations. We introduced a vacuum of $\sim$25 $\textrm{Å}$between
the two layers of ScS$_{2}$ to avoid any interaction with the periodic
image in that direction in order to realize the monolayer. A 3$\times$3$\times$1
supercell was considered to study the adsorption of the metal atoms
so that the change in the binding energy and average open circuit
voltage (OCV) with the concentration of adatoms can be studied in
detail. For geometry optimization of the absorbed system 6$\times$6$\times$1
$k$-points were considered for the Brillouin zone integration within
the Monkhorst\textminus Pack scheme\cite{29}. To study the diffusion
of a single metal atom on the ScS$_{2}$ surface we employed the climbing-image
nudged elastic band (CI-NEB) method\cite{30}. In this method the
images initially defined by the linear interpolation between the two
minima are relaxed to define a minimum energy path (MEP). DFT computes
the forces both on the images and the elastic band. The MEP is defined
when the component of the force not pointing along the path direction
defined by the images is zero. We considered the force convergence
criteria for the CI-NEB calculations as 0.01 eV/Å.

\section{Results and Discussions: }

\subsection{Structure of ScS$_{2}$ Monolayer:}

\begin{figure}
\begin{centering}
\includegraphics[width=0.8\textwidth,keepaspectratio]{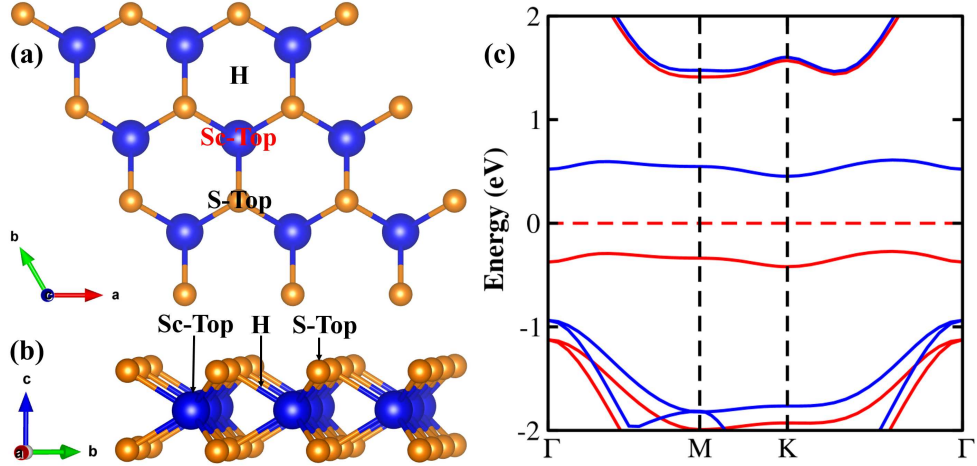}
\par\end{centering}
\caption{\label{Figure1}(a) Top-view of the optimized 3X3X1 supercell of ScS$_{2}$
monolayer marked with possible adsorption sites (b) side-view of the
optimized 3X3X1 supercell of ScS$_{2}$ monolayer marked with possible
adsorption sites. Blue and orange balls represent Sc and S atoms respectively.
(c) Calculated electronic band-structure of H-ScS$_{2}$ monolayer.
Spin-up states are marked with red and spin-down states are marked
with blue colour. }
\end{figure}

The optimized structure of monolayer ScS$_{2}$ is shown in the Figure
\ref{Figure1}(a) and (b). Here, the structure is two-dimensional
and consists of one ScS$_{2}$ sheet oriented along the (0, 0, 1)
direction consisting one Sc-layer sandwiched between two S-layers.
Due to the hexagonal symmetry, each Sc atom occupies the trigonal
prismatic centre position by sharing bonds to six equivalent S atoms
in the 6-coordinated geometry of the trigonal prism. We obtained the
$Sc-S$ bond length and the $\angle S-Sc-S$ bond angle as 2.57$\textrm{Å}$
and $\sim$95$\lyxmathsym{\textdegree}$, respectively. The optimized
lattice parameter determined from our calculations (3.79$\textrm{Å}$)
is in good agreement with the previous studies\cite{21,22}. The electronic
band-stucture of the optimized H-ScS$_{2}$ is shown in the Figure
\ref{Figure1}(c). We found an indirect bandgap of 0.73 eV which also
matches with the previous predictions of 0.74 eV\cite{21} and 0.72
eV\cite{22}. As the experimental report on the synthesis of single
layer ScS$_{2}$ is not available yet to best of our knowledge, the
next important step is to theoretically explore the possibility of
exfoliation of monolayer of ScS$_{2}$ from bulk. For this we calculated
the exfoliation energy per atom which is basically the average energy
per atom required to remove a layer from it's layered bulk counterpart,
using the formula:
\begin{equation}
E_{exf}=\frac{E_{mono}}{N_{mono}}-\frac{E_{bulk}}{N_{bulk}}.
\end{equation}
Here, E$_{mono}$ and E$_{bulk}$ are the total energies of the monolayer
and the bulk materials, respectively, and N$_{mono}$ and N$_{bulk}$
are the number of atoms in the monolayer and in the bulk, respectively.
We obtained the value as 150 meV/atom for H-ScS$_{2}$ and reporting
here for the first time to best of our knowledge. It is well within
the desirable limit of 200 meV/atom\cite{31}, although materials
having exfoliation energy as high as 260 meV/atom like Ca$_{2}$N\cite{32}
and $\sim$600 meV/atom for silicene and GeO, have already been
synthesized experimentally. Therefore, it can very well be predicted
that the single layer of ScS$_{2}$ can be easily exfoliated from
its bulk counterpart. The stability of the H-ScS$_{2}$ monolayer
has already been shown in the previous papers\cite{21,33}.

\subsection{Adsorption of the Cations: }

\begin{figure}
\begin{centering}
\includegraphics[width=0.6\textwidth]{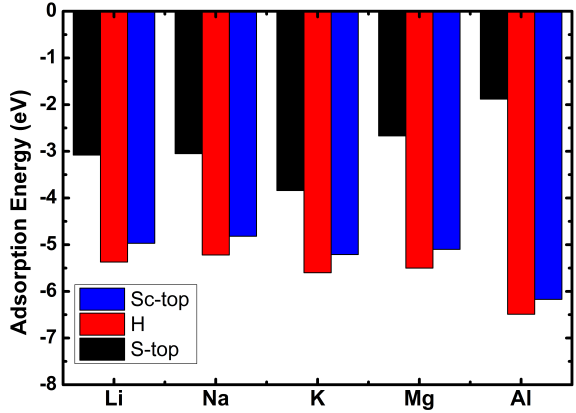}
\par\end{centering}
\caption{\label{Figure2}Adsorption energy of different cations on the ScS$_{2}$
monolayer surface at different adsorption sites}
\end{figure}

Adsorption energy of metal atoms on the electrode material is a crucial
parameter in determining the electrochemical performance of cathode
or anode. For cathode, a relatively high adsorption energy is expected.
To study the metal atoms adsorption on the ScS$_{2}$ surface we considered
one adatom in a 3$\times$3$\times$1 supercell which is big enough
to avoid any interactions between the adsorbates. There are three
possible adsorption sites based on the symmetry and chemical environment,
namely: (a) H site, (b) Sc-top site and (c) S-top site as marked in
the Figure\ref{Figure1}(a) and (b). The adsorption energies of the
cations on these sites were calculated by the formula: 
\begin{equation}
E_{ads}=E_{ScS_{2}M_{n}}-E_{ScS_{2}}-nE_{M}
\end{equation}
where, E$_{ScS_{2}M_{n}}$ is the total energy of the metal atom(s)
adsorbed ScS$_{2}$ monolayer, E$_{ScS_{2}}$ is the total energy
of the pristine ScS$_{2}$ monolayer, E$_{M}$ is the energy per atom
of the metal cation calculated from the corresponding bulk phase and
$n$ is the number of metal atoms adsorbed per ScS$_{2}$ formula
unit. We considered the bulk bcc (body-centred cubic) structure for
Li, Na and K and primitive hexagonal for Mg and fcc (face-centred
cubic) for Al to calculate E$_{M}$. The calculated adsorption energies
of the cations on each of the adsorption sites are shown in the Figure
\ref{Figure2}. Negative values of the adsorption energy indicate
the exothermic interactions between the adsorbent and adsorbates and
also signify effective adsorption of all the metal atoms on the ScS$_{2}$
monolayer which is not the general case for TMDCs\cite{4}. From the
Figure \ref{Figure2} it can be seen that H site is the most favorable
adsorption site for all the metal atoms, followed by Sc-top site and
S-top site. This can be explained from the fact that generally chemisorbed
atoms prefer the high-coordination sites\cite{34}. The detail explanation
is given later on. On comparing the adsorption energies of Li (-5.37
eV), Na (-5.22 eV) and K (-5.6 eV) on ScS$_{2}$ to the other cathode
materials like ScO$_{2}$ (-3.18 eV for Li, -2.87 eV for Na, -9.18
for K)\cite{17}, MnO$_{2}$ (-2.54 for Li, -2.38 for Na)\cite{9},
CoO$_{2}$ (-3.60 for Li, -3.40 for Na)\cite{9}, NiO$_{2}$ (-2.17
for Li, -2.56 for Na)\cite{9}, NbS$_{2}$ (-2.03 for Li, -1.80 for
Na)\cite{35} etc. it is found that ScS$_{2}$ attracts the alkali
metal atoms more strongly which is desirable for cathode materials.
It should be noted here that in contrast to our case, for ScO$_{2}$
monolayer there is a hexagonal to trigonal phase transition at very
low concentration of Li and K adsorption. In fact for the whole range
of concentration of the adsorbed metal atoms, no significant structural
changes were observed for ScS$_{2}$ monolayer which indicates possible
electrochemical reversibility of the material. Also for Mg (-5.5 eV)
and Al (-6.49 eV), ScS$_{2}$ poses higher adsorption energy than
other materials like MoS$_{2}$ nanoribbon (-4.85 eV for Mg)\cite{20,4}.
We further calculated the binding energy of the metal atoms by the
formula:
\begin{equation}
E_{binding}=[E_{ScS_{2}M_{n}}-E_{ScS_{2}}-nE_{M}]/n
\end{equation}
which basically is same as Equation 3.2 for n=1. 

\begin{figure}
\centering{}\includegraphics[width=0.5\textwidth,keepaspectratio]{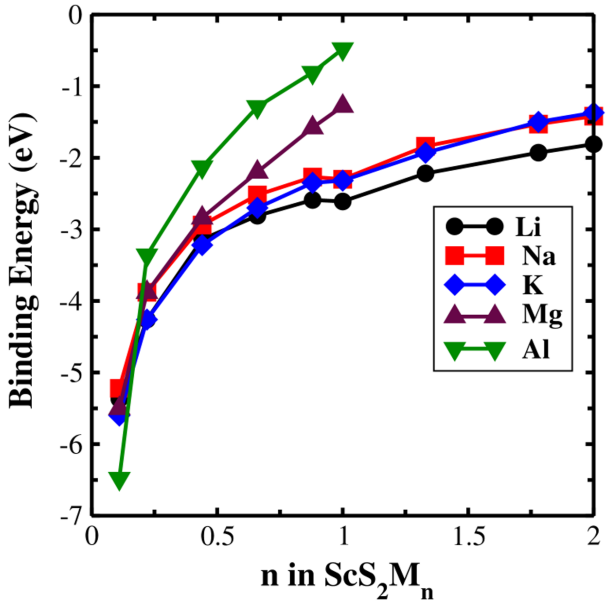}\caption{\label{Figure3}Calculated binding energies for different metals on
the ScS$_{2}$ surface as a function of metal atom concentration }
\end{figure}

Figure \ref{Figure3} shows the binding energy as function of ion
concentration (M$_{n}$ScS$_{2}$) in the 3$\times$3$\times$1 supercell
of ScS$_{2}$. The binding energy remains negative for the whole range
indicating the metal atoms will prefer to be absorbed on the host
material even in higher concentration rather than forming metal cluster
themselves\cite{2}. However, the adsorption energy decays with the
increase in the ion concentration due the electrostatic repulsive
forces among the ions. As K has the highest ionic radius among the
alkali metals studied, the increase in the ion concentration for K
results in more electrostatic repulsion, and hence, higher reduction
in the binding energy than others.

\begin{table}
\caption{\label{Table1}Metal atoms adsorption on the H-site of ScS$_{2}$
monolayer: Net Bader charge on different atoms (q$_{M}$: Metals atoms,
q$_{S}$: each of the neighboring three S atoms$^{a}$) and the distances
of the metal atoms adsorbed in the H site from the neighboring three
Sc atoms (D$_{M-Sc}$) and three S atoms (D$_{M-S}$) (Figure \ref{Figure1})}

\begin{centering}
\begin{tabular}{ccccc}
\hline 
Metal Atom (M) & q$_{M}$ (e) & q$_{S}$ (e) & D$_{M-Sc}$ ($\textrm{Å}$) & D$_{M-S}$ ($\textrm{Å}$)\tabularnewline
\hline 
Li & 0.86 & -1.14 & 3.49 & 2.35\tabularnewline
Na & 0.78 & -0.92 & 3.89 & 2.67\tabularnewline
K & 0.81 & -0.89 & 4.35 & 3.04\tabularnewline
Mg & 1.27 & -1.03 & 3.38 & 2.35\tabularnewline
Al & 1.47 & -0.95 & 3.16 & 2.16\tabularnewline
\hline 
\end{tabular}
\par\end{centering}
$^{a}$The net Bader charge on each of the three neighboring S atoms
of the H site in the pristine ScS$_{2}$ is -0.74 e
\end{table}

Table \ref{Table1} shows the different important parameters related
to the adsorption of metal adatoms on the ScS$_{2}$ surface. As the
differences in the electronegativity between S and the metal atoms
are quite high than those of the Sc and the metal atoms, the adsorbates
are more closely and strongly bonded to the three neighboring S atoms
than the Sc atoms as evidenced by the less metal$-$S distances (D$_{M-S}$)
than the corresponding metal$-$Sc distances (D$_{M-Sc}$) given in
the Table \ref{Table1}. Due to the stronger electronegitivity of
S than that of the metal atoms, all the cations act as electron donors.
From the Bader charge analysis as given the Table \ref{Table1} it
can be seen that Li, Na, K, Mg and Al donates 0.86e, 0.78e, 0.81e,
1.27e and 1.47e, respectively, to the adsorbent system. On comparing
the electron transfer of the Li, Na and K atoms on the FeSe surface
and the charge transfer to the Cl atoms in their respective chloride
compounds (LiCl, NaCl and KCl)\cite{36}, we found that these atoms
on the ScS$_{2}$ monolayer surface can safely be taken as fully ionized,
i.e, in the charge state of +1. We also found that, the Mg atom adsorbed
on the ScS$_{2}$ surface might not get fully ionized (but will be
very close to it) as the earlier comparison as well as the study on
VS$_{2}$ surface\cite{4} reveals a charge transfer about 1.5e as
equivalent to full ionization or charge state of +2. However, Al atom
adsorbed in our case, is nowhere close to full ionization (charge
state of +3), which can be understood from its relatively higher elctronegativity
(i.e., lower difference of electronegativity with S) among all the
metal atoms considered. As expected, all these electrons from the
metal atoms has been transferred to the three neighboring S atoms
of the H site as evidenced by the increase of the net Bader charges
of these atoms from the pristine ScS$_{2}$ (Table \ref{Table1}).
Practically metal atoms do not interact directly with Sc atoms but
only interact with S atoms which also explains the probable reason
of H site being the most favorable adsorption site as in this configuration
the metal atoms can maximize the S-coordination number and effectively
interact with more no of S atoms than the other configurations.

\begin{figure}
\begin{centering}
\includegraphics[width=1\textwidth,keepaspectratio]{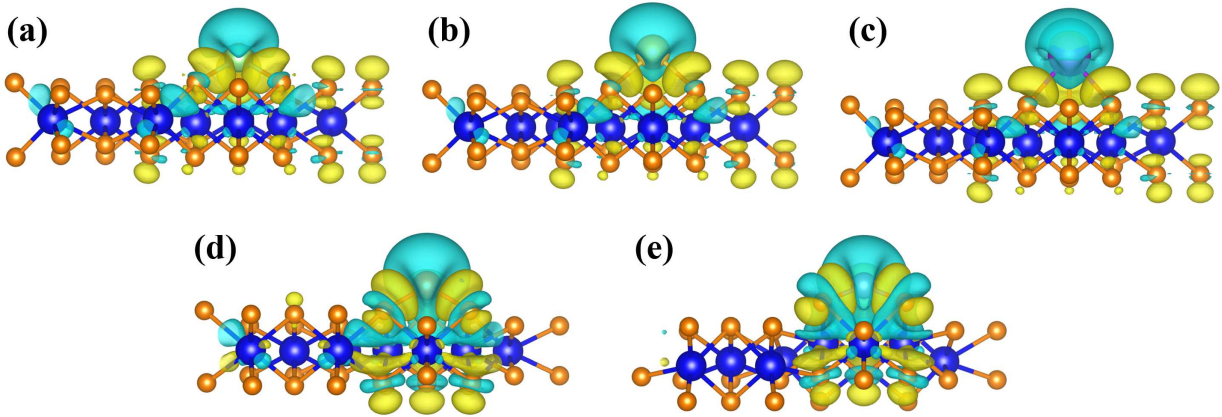}
\par\end{centering}
\caption{\label{Figure4}Plot of differential charge density of (a) Li, (b)
Na, (c) K, (d) Mg, and (e) Al atom adsorbed on the ScS$_{2}$ monolayer.
The iso-surface value is taken as 0.0075 e/$\textrm{Å}^{3}$. Yellow
and cyan surfaces represent the charge accumulation and depletion
region respectively.}
 
\end{figure}

To visualize the nature of the charge transfer process we calculated
the differential charge density defined by the formula:
\begin{equation}
\rho_{diff}=\rho_{ScS_{2}+M}(r)-\rho_{ScS_{2}}(r)-\rho_{M}(r)
\end{equation}
where $\rho_{ScS_{2}+M}(r)$ is the charge density distribution of
the combined system of metal atom absorbed on ScS$_{2}$ monolayer,
$\rho_{ScS_{2}}(r)$ is the charge density of the ScS$_{2}$ monolayer,
and $\rho_{M}(r)$ represents the charge density of the metal atom.
Figure \ref{Figure4} depicts the visualization of differential charge
density plot of the metal atoms adsorption on the ScS$_{2}$ surface.
From the figure it can be seen that for alkali metal atoms (Li, Na,
K) there are distinct positive (charge depletion) and negative region
(charge accumulation) of charge densities located on metal atoms and
on neighboring S atoms, respectively, signifying the ionic characteristics
of the metal-S bonds which is in accordance to the Bader charge analysis
discussed above. For Mg, the situation is more or less same except
a small shift of the charge accumulation regions from the vicinity
of S atoms towards the Mg atom which may be due to some covalent characteristics
of the Mg-S bonds, as discussed earlier too, that the Mg atom adsorbed
on the ScS$_{2}$ surface is not in the full ionization state but
very close to it. However, in case of Al adsorbtion, the covalent
component of the Al-S bonds can be clearly seen from the figure. There
are charge accumulation region between the adsorbed Al atom and the
S atoms and also alternative depletion and accumulation regions. In
accordance to the Bader charge analysis, differential charge density
plot also suggest that Al-S bonds are not fully ionic bond like others
but, also have major covalent components. It should be also noted
that for none of the cases any significant charge accumulation or
depletion happens on the Sc atoms which supports the fact that metal
atoms interact only with S atoms and not with Sc atoms.

\subsection{Voltage Profile and Theoretical Capacity:}

Open circuit voltage (OCV) and theoretical storage capacity are two
crucial parameters to determine the performance of electrode materials.
Optimization of these two parameters are also important for designing
cathode materials. To study OCV and specific capacity we used a 3$\times$3$\times$1
supercell of ScS$_{2}$ and considered the adsorption on both sides.
The charge-discharge process of ScS$_{2}$ single layer can be described
by the following half-cell reaction \textit{vs }M/M$^{x+}$ as:
\begin{equation}
ScS_{2}+nM^{x+}+nxe^{-}\leftrightarrow ScS_{2}M_{n}
\end{equation}
 Now, depending on this reaction the OCV can be calculated from the
difference in total energy of the monolayer before and after adsorption
of the metal atoms. Neglecting the volume term ($P\nabla V$) and
entropy term ($T\nabla S$) the average OCV (V$_{avg}$) can be expressed
as\cite{37}:
\begin{equation}
V_{avg}=(E_{ScS_{2}}+nE_{M}-E_{ScS_{2}M_{n}})/nxe
\end{equation}
where $x$ is the charge state of fully ionized cations in the electrolyte,
i.e., $x=1$ for Li, Na and K; $x=2$, for Mg and $x=3$ for Al. 

\begin{figure}
\centering{}\includegraphics[width=1\textwidth,keepaspectratio]{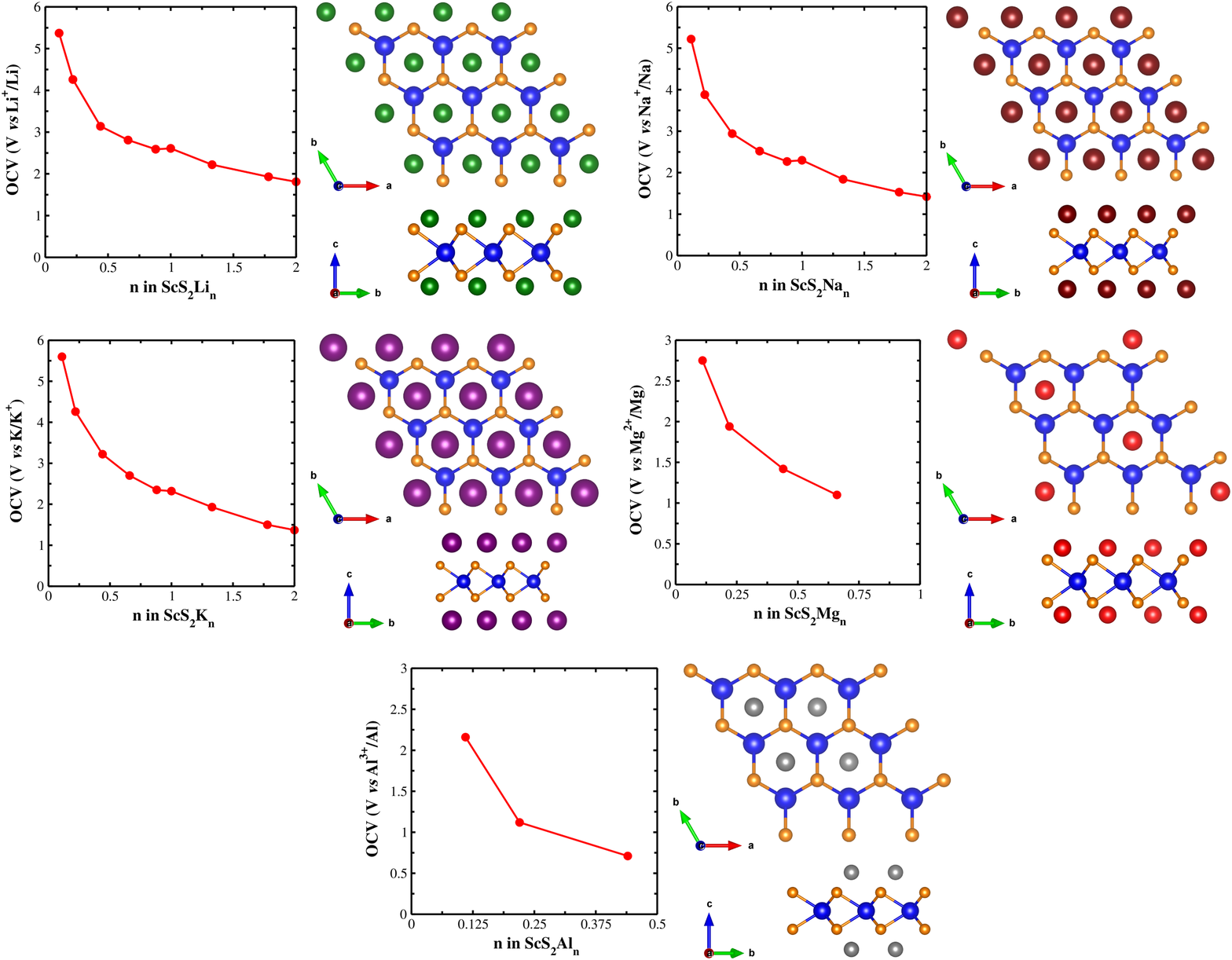}\caption{\label{Figure6}Calculated average cell voltage profile of ScS$_{2}$
monolayer and structre with highest no of metal ions absorbed for 
Li, Na (first row), K, Mg (second row), and Al (third row) adsorption}
\end{figure}

The calculated average OCV up to fully covered double side adsorption
for alkali ions (M$_{2}$ScS$_{2}$; M = Li, Na, K) and partially
covered double side adsorption for Mg and Al ions (MScS$_{2}$; M
= 0.66 for Mg, 0.44 for Al) has been shown in the Figure \ref{Figure6}.
The corresponding maximum capacities for different cations have been
calculated using the equation:
\begin{equation}
C=nxF/M_{ScS_{2}}
\end{equation}
 where, $F$ is the Farady constant (26801 mA h $mol^{-1}$) and $M_{ScS_{2}}$
is the molecular weight of ScS$_{2}$ (109.09 g $mol^{-1}$). The
maximum theoretical capacities for the alkali metal ions have been
calculated to be 491.36 mA h g$^{-1}$. At the highest capacity (Li$_{2}$ScS$_{2}$),
the Li-cell voltage is 1.81 V whereas at the concentration of LiScS$_{2},$
which gives the capacity of 245.68 mA h g$^{-1}$, the cell voltage
remains as high as 2.61 V. For lithiation in ScO$_{2}$, it has been
reported to have average OCV as 4.15 V but at the highest capacity
of 348 mA h g$^{-1}$ only, and beyond that voltage goes down below
1 V. Also, at the same time a hexagonal to trigonal phase transition\cite{17}
takes place which may affects the electrochemical reversibility of
the material, as discussed earlier. In fact Liu \textit{et al. }reported
the same problem of ScO$_{2}$ for Na and K, as well. ScS$_{2}$ as
LIB cathode is also found to perform well in term of average OCV than
MnO$_{2}$ (V$_{avg}$= 1.11 V at 308.5 mA h g$^{-1}$ and 0.98 V
at 617 mA h g$^{-1}$)\cite{9} and similarly to CoO$_{2}$\cite{9}.
On the other hand, V$_{2}$O$_{5}$ monolayer can deliver specific
capacity of 147.40 mA h g$^{-1}$ (LiV$_{2}$O$_{5}$) and 294.79
mA h g$^{-1}$ (Li$_{2}$ V$_{2}$O$_{5}$) only for Li at an average
OCV of 2.43 V and 2.06 V, respectively\cite{18}. Thus, taking into
account both the parameters, i.e., OCV and specific capacity, ScS$_{2}$
monolayer can be a preferred choice over V$_{2}$O$_{5}$ monolayer
also. In the study of sodiation of NbS$_{2}$\cite{19}, Liao \textit{et
al. }reported the DFT results for OCV calculations which are found
to be in good agreement with experimental results. Inspired from this,
we also compared the performance of ScS$_{2}$ as LIB cathode with
the recent experimentally studied other sulfide materials like 1T-VS$_{2}$
or 1T-MoS$_{2}$\cite{38,39} and found ScS$_{2}$ to be a better
candidate. 

Like Li, in case of Na also, ScS$_{2}$ performs better than ScO$_{2}$\cite{17}
if the factor of higher capacity and electrochemical reversibility
are taken into account as it can deliver the specific capacity of
245.68 mA h g$^{-1}$ (NaScS$_{2}$) at 2.3 V and 491.36 mA h g$^{-1}$
(Na$_{2}$ScS$_{2}$) at 1.42 V without any significant structural
change, as shown in Figure \ref{Figure6}. It has also found to be
a better choice as NIB cathode in term of OCV over oxide layered materials
like MnO$_{2}$, CoO$_{2}$ and NiO$_{2}$\cite{9} which has average
OCV of 0.48 V, 0.88 V, and 1.02 V, respectively. V$_{2}$O$_{5}$
monolayer on the other hand has found to deliver the specific capacity
of 147.40 mA h g$^{-1}$ (NaV$_{2}$O$_{5}$) at 2.54 V average OCV
and 294.79 mA h g$^{-1}$ (Na$_{2}$ V$_{2}$O$_{5}$) at 1.86 V average
OCV which when compared to ScS$_{2}$ monolayer it can be seen that
for higher capacity at moderate high OCV the later one can be a preferred
choice. Recently studied sulfide materials as NIB cathode like NbS$_{2}$\cite{19}and
Cu$_{2}$S\cite{40} have been found to have voltage window between
2.75 V to 1 V and specific capacity of 170.66 mA h g$^{-1}$ and OCV
range of 2.6 V to 0.4 V and specific capacity of 294 mA h g$^{-1}$,
respectively, which in every aspect perform poor to ScS$_{2}$ monolayer. 

As a K-ion battery cathode material, ScS$_{2}$ is also found to perform
well, delivering the specific capacity of 245.68 mA h g$^{-1}$ (KScS$_{2}$)
at an average OCV of 2.32 V and 491.36 mA h g$^{-1}$ (K$_{2}$ScS$_{2}$)
at 1.37 V. In comparison, V$_{2}$O$_{5}$ monolayer can only deliver
the specific capacity of 147.40 mA h g$^{-1}$ at slightly higher
2.77 V average OCV\cite{18}. Mathew \textit{et al. }have studied
the insertion of K-ions in the amorphous iron phosphate and reported
a voltage range of $3.5-1.5$ V with highest capacity around 150 mA
h g$^{-1}$\cite{41}. To compare with their data we also calculated
the step potential of K adsorption on ScS$_{2}$ using the following
equation: 
\begin{equation}
V_{step}=\frac{E_{ScS_{2}M_{n_{1}}}+(n_{2}-n_{1})E_{M}-E_{ScS_{2}M_{n_{2}}}}{(n_{2}-n_{1})xe}
\end{equation}
where, n$_{1}$ and n$_{2}$ (n$_{2}$>n$_{1}$) are the metal ion 
concentrations in one ScS$_{2}$ formula unit in two successive steps.
It gives the voltage range of $5.6-2.05$ V with capacity of 245.68
mA h g$^{-1}$at an average OCV of 2.32 V. MoN$_{2}$ has been reported
to have a theoretical capacity of 432 mA h g$^{-1}$ at an average
OCV of 1.11 V which is again not as good as ScS$_{2}$ in terms of
both average OCV and specific capacity. Other than these alkali metal
ions, 2D ScS$_{2}$ has also been found to perform well as cathode
of multivalent cations like Mg and Al in terms of OCV and specific
capacity. For Mg, it can deliver maximum of 324.29 mA h g$^{-1}$
(Mg$_{0.66}$ScS$_{2}$) theoretical capacity at an average OCV of
1.1 V beyond which the step potential becomes negative. For comparison,
V$_{2}$O$_{5}$ monolayer can deliver 294.79 mA h g$^{-1}$ theoretical
capacity at 1.02 V average OCV and its bulk counterpart delivers even
less specific capacity of 260 mA h g$^{-1}$ at OCV 1.83 V\cite{18}.
On the other hand one of the best known Mg-ion cathode material Mo$_{6}$S$_{8}$
is known to produce charge-storage capacity of 110 mA h g$^{-1}$
only at an operating voltage $\sim$1.2 V\cite{42}. CuS-- studied
for the same purpose show an average storage capacity $\sim$300 mA
h g$^{-1}$ in the voltage range of 2.4 V to 0 V which can be compared
to our step voltage potential range of 2.75 V to 0.44 V\cite{42}
which is better for the obvious reason. Next, for Al we found ScS$_{2}$
monolayer to deliver same theoretical specific capacity of 324.29
mA h g$^{-1}$ (Al$_{0.44}$ScS$_{2}$) like Mg at an average OCV
of 0.71 V. The step potential window in this case in the range of
2.16 V -- 0.31 V. To give a context to our findings we compared it
with recently studied layered sulfide materials like TiS$_{2}$\cite{43}
which experimentally gives a voltage range of $\sim$1.3 V$-$0.2
V and specific capacity of $\sim$50 mA h g$^{-1}$ only, MoS$_{2}$\cite{44}
delivers specific capacity of 253.6 mA h g$^{-1}$ with a voltage
between 2 V--0.5 V. ScS$_{2}$ also predicted to theoretically perform
better than other TMDCs that has been studied as Al-ion battery cathode
materials like TiO$_{2}$ (1.1 V$-$0.4 V, 75 mA h g$^{-1}$)\cite{45},
VO$_{2}$ (0.9 V$-$0.01 V, 116 mA h g$^{-1}$)\cite{46}.

Most of the cathode materials studied so far for metal ion batteries
are not able to deliver large specific capacity while holding high
OCV. In this context, ScS$_{2}$ can be an interesting candidate for
cathode material as it maintains a good balance between high operating
voltage window and the amount of ions it can hold i.e., the specific
capacity. 

\subsection{Diffusion Kinetics: }

\begin{figure*}[h]
\centering{}\includegraphics[width=1\textwidth,keepaspectratio]{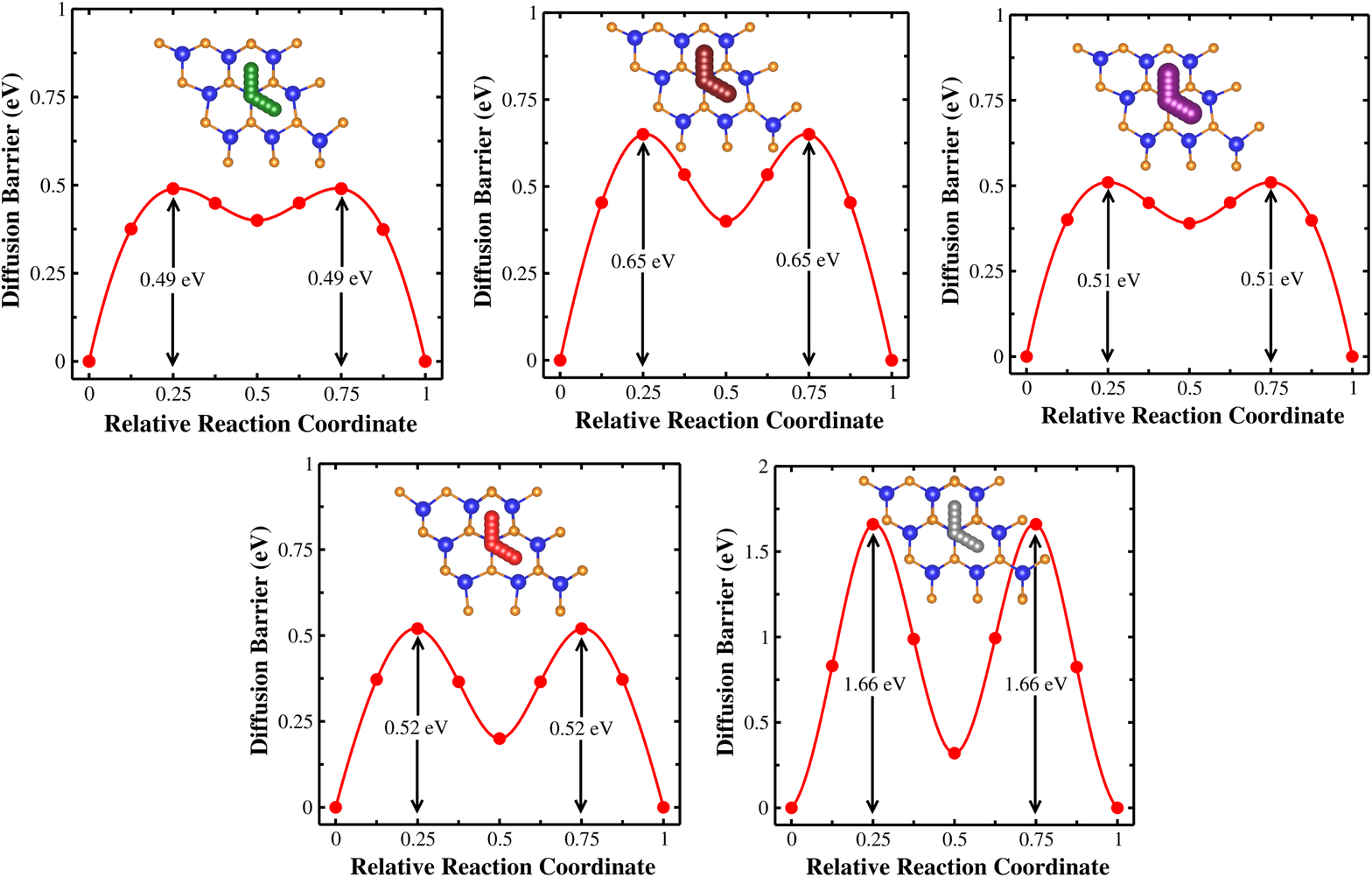}\caption{\label{Figure7}Diffusion barrier profile for the (i) Li, (ii) Na,
(iii) K (iv) Mg and (v) Al atoms on the ScS$_{2}$ monolayer surface}
\end{figure*}

The diffusion kinetics of the ions in the cathode is directly related
to the rate of the charge/discharge process of the rechargeable batteries
and hence it is a very crucial parameter to study. For this purpose
we employed the CI-NEB method\cite{30} as discussed in the computational
details section to find the diffusion barrier height (E$_{B}$) and
the minimum energy path. From the study of the adsorption of the adatoms
on the ScS$_{2}$ surface, we have seen that H is the most energetically
favorable position followed by Sc-top and S-top sites. Accordingly
we have predesigned two diffusion pathways, (i) cations move from
one H site to the adjacent H site through a Sc-top site and (ii) cations
move from one H site to the another H site in a linear path. Here
we did not consider a third possibility of cation diffusion along
the path H to H \textit{via }S-top site because of the fact that the
adsorption energies at the S-top site are significantly less than
the other two sites indicating a higher diffusion barrier which has
also been seen for other 2D materials\cite{4}. We considered seven
images between the two H sites. After optimization we found that the
predesigned path (ii) actually obtained the same pathway as (i), which
indicates path (i) to be the probable minimum energy path for cation
diffusion on the ScS$_{2}$ surface. The optimized diffusion pathway
for the five adatoms are shown in the Figure\ref{Figure7}. In this
$H-Sc-top-H\text{ pathway}$ one metastable state i.e., the local
energy minima was found at the Sc-top site for all of the adsorbed
cations. Also two saddle points are observed in the middle of the
H$-$Sc-top bridges. The diffusion energy barriers for Li, Na, K,
Mg and Al are found to be 0.49 eV, 0.65 eV, 0.51 eV, 0.52 eV and 1.66
eV respectively as shown in the Figure\ref{Figure7}. For the alkali
atoms the diffusion energy barriers are in the order of E$_{B-Li}$<E$_{B-K}$<E$_{Na}$,
which is also the case for V$_{2}$O$_{5}$\cite{18} {[}Ref: V2O5{]}.
The barrier height for Li-atom on the ScS$_{2}$ surface is less than
the previously reported 2D materials like MoN$_{2}$\cite{2} and
TiO$_{2}$\cite{47} for which the respective heights are 0.78eV and
0.65 eV and also from some of the bulk anodes like Si (E$_{B}$=0.62
eV)\cite{48} and cathodes like LiFePO$_{4}$ (E$_{B}$=0.60 eV)\cite{48a,48b,48c}
and Li$_{2}$FeSiO$_{4}$ (E$_{B}$=0.91 eV)\cite{49}. In case of
Na diffusion, the migration barrier is slightly higher than Li on
ScS$_{2}$ but it's still better than many other electrodes like V$_{2}$O$_{5}$\cite{18},
Na$_{0.33}$Si$_{24}$\cite{50}, Si\cite{48}, Ge\cite{48} etc.,
the respective barrier heights are 1.17 eV, 0.68eV, 1.08 eV and 0.78
eV. For K-atom migration on the monolayer of ScS$_{2}$, the barrier
height is almost equal to that of the MoN$_{2}$ surface (E$_{B}$=0.49
eV)\cite{2} and better than that of the bulk-V$_{2}$O$_{5}$ (E$_{B}=$1.66
eV)\cite{18}. We found the Mg diffusion on the ScS$_{2}$ more facile
than on the surface of TiS$_{2}$ (E$_{B}$=0.55 eV)\cite{52}, V$_{2}$O$_{5}$
(E$_{B}$=1.36 eV)\cite{18}, and FeSe (E$_{B}$=0.85 eV)\cite{36}.
Our result is also comparable to the migration energy barrier for
Mg on the zigzag MoS$_{2}$ nanoribbon surface (E$_{B}$=0.48 eV)\cite{20},
which has been studied as a promising cathode material for Mg-ion
battery. Al has the highest diffusion energy barrier among the five
cations we studied. Because of the unavailability of the theoretical
results on the same for the cathode materials of the Al-ion battery,
it's hard to compare the performance of our material directly to other
materials in this aspect. However, the migration barrier for K atom
in the bulk of V$_{2}$O$_{5}$\cite{18} is same as of Al on ScS$_{2}$
surface. From the previous discussion on the adsorption of adatoms
on the monolayer of ScS$_{2}$ we have seen that the Al is not fully
ionized and the Al-S bonding has some covalent character unlike other
atoms. The adatoms sitting on the H-site have the S coordination number
as three. Now in the course of migration from H-site to Sc-top site,
adatoms cross the saddle point which is found to be positioned in
the middle of the H$-$Sc-top bridge which is two-coordinated w.r.t.
S atoms. As the covalent bonds are the directional bonds, the difference
in coordination number between any two adsorption sites (in this case,
between the ground state and the transition state) results in higher
change in energy (in this case, the diffusion barrier) than in the
case of ionic bonds which are non-directional\cite{34,53} in nature.
The fact that the difference between the adsorption energies on S-top
site (one-coordinated w.r.t. S atom) and H-site for Al atom is the
highest among all also is in accordance with this argument. So, it
may be the reason for Al to have highest diffusion barrier among the
studied cations.

From the above discussion on the diffusion kinetics of the five cations
on the ScS$_{2}$ surface and the comparison with other materials,
we can infer that Li, Na, K and Mg can migrate with ease indicating
a good charge/discharge performance. However, for Al, the diffusion
barrier is high and needs further efforts to minimize it. For example,
in case of Li, Guo et al.\cite{54} shown that the diffusion barrier
on phosphorene sheet can be significantly reduced after introducing
specific vacancy defects. 

\bibliography{refs}

\end{document}